\documentclass[journal=nalefd,manuscript=letter]{achemso}
\setkeys{acs}{articletitle = false}
\usepackage{xcolor,graphicx,mathtools,braket,natbib}
\usepackage[colorlinks=true, allcolors=black]{hyperref}
\usepackage[caption=false]{subfig}
\usepackage[T1]{fontenc}
\usepackage[font=small]{caption}

	\author{D.L.~Hurst}
	\affiliation{Department of Physics and Astronomy, University of Sheffield, Hounsfield Road, Sheffield, S3 7RH, United Kingdom}\author{D.M.~Price}
	\affiliation{Department of Physics and Astronomy, University of Sheffield, Hounsfield Road, Sheffield, S3 7RH, United Kingdom}
	\author{C.~Bentham}
	\affiliation{Department of Physics and Astronomy, University of Sheffield, Hounsfield Road, Sheffield, S3 7RH, United Kingdom}
	\author{M.N.~Makhonin}
	\affiliation{Department of Physics and Astronomy, University of Sheffield, Hounsfield Road, Sheffield, S3 7RH, United Kingdom}
	\author{B.~Royall}
	\affiliation{Department of Physics and Astronomy, University of Sheffield, Hounsfield Road, Sheffield, S3 7RH, United Kingdom}
	\author{E.~Clarke}
	\affiliation{EPSRC National Epitaxy Facility, Department of Electronic and Electrical Engineering, University of Sheffield, Sheffield S1 3JD, United Kingdom}
	\author{P.~Kok}
	\affiliation{Department of Physics and Astronomy, University of Sheffield, Hounsfield Road, Sheffield, S3 7RH, United Kingdom}
	\author{L.R.~Wilson}
	\affiliation{Department of Physics and Astronomy, University of Sheffield, Hounsfield Road, Sheffield, S3 7RH, United Kingdom}
	\author{M.S.~Skolnick}
	\affiliation{Department of Physics and Astronomy, University of Sheffield, Hounsfield Road, Sheffield, S3 7RH, United Kingdom}
	\author{A.M.~Fox} 
	\email{mark.fox@sheffield.ac.uk}
	\affiliation{Department of Physics and Astronomy, University of Sheffield, Hounsfield Road, Sheffield, S3 7RH, United Kingdom}
	\date{\today}

\title {Non-Reciprocal Transmission and Reflection of a Chirally-Coupled Quantum Dot}


\begin{document}
	
		\begin{abstract}
		\noindent We report strongly non-reciprocal behaviour for quantum dot exciton spins coupled to nano-photonic waveguides under resonant laser excitation. A clear dependence of the transmission spectrum on the propagation direction is found for a chirally-coupled quantum dot, with spin up and spin down exciton spins coupling to the left and right propagation directions respectively. The reflection signal shows an opposite trend to the transmission, which a numerical model indicates is due to direction-selective saturation of the quantum dot. The chiral spin-photon interface we demonstrate breaks reciprocity of the system and opens the way to spin-based quantum optical components such as optical diodes and circulators in a chip-based solid-state environment. 
	\end{abstract}
	
	\noindent
	The deterministic coupling of a two-level system to a one-dimensional waveguide provides a near-ideal platform for demonstrating quantum-optical effects such as single-photon nonlinearities \cite{cite-key}. A key parameter for such ``1-D atoms'' is the $\beta$-factor, which quantifies the relative coupling to the waveguide compared to other optical modes. In the limit of $\beta\rightarrow1$ and with no decoherence, the scattering of a single photon results in its complete reflection, leading to a 100\% dip in the transmission spectrum \cite{PhysRevA.82.063821}. Such effects have been observed in a variety of systems, notably semiconductor quantum dots (QDs) coupled to photonic crystal waveguides \cite{Javadi_2015,Hallett:18} and SiV or GeV centres coupled to nanobeams \cite{Sipahigil847,Bhaskar,PhysRevApplied.8.024026}, with transmission dips as large as 60\% now reported \cite{Thyrrestrup_2017}.
	
	The recent discovery of non-reciprocal coupling between dipole emitters and nano-photonic structures \cite{Junge-2013,Petersen,Rodriguez,Sollner,Lodahl-review,2040-8986-19-4-045001,ncomms,Kuipers} adds a new dimension to the system. These chiral effects arise from the spin-orbit interaction of light \cite{spin-orbit-light} and lead to directionality in the $\beta$-factor, with circular dipoles of opposite sense coupling to modes propagating in opposite directions. The result of a transmission-type experiment on a chirally-coupled emitter has to be different to the non-chiral case, as the emitter does not couple to the backward propagating mode and hence reflection is not possible. In the coherent, single-photon limit, with $\beta \rightarrow 1$, light is transmitted with 100\% probability and so the transmission dip on resonance is now expected to be negligibly small.

	The ideal behaviour is hard to observe in practice: the $\beta$-factor is never perfect and dephasing is always present to some extent. Moreover, the directional coupling efficiency is less than unity. In these non-ideal conditions, the behaviour is expected to lie somewhere between the limits of perfect reflection and perfect transmission for the non-chiral and chiral cases respectively. 
	In this paper, we present experimental data on a single QD chirally coupled to a nanobeam waveguide and then use a theoretical model to describe the system. The key finding is the observation of a spin-dependent dip in the transmission, which depends strongly on the direction of propagation, thereby breaking reciprocity. 
	We also present experimental data on directional spin-dependent reflectivity, where, unexpectedly, the more weakly coupled dipole gives the larger signal. The theoretical modelling shows that this counter-intuitive behaviour is caused by the increased saturation of the more strongly coupled QD spin at the power levels used in the experiment.
	The use of a  semiconductor emitter fully integrated into a single-mode nanophotonic waveguide leads to a much larger overall $\beta$-factor than used in previous work on non-reciprocal transmission for cold atoms coupled to a nanofiber~\cite{PhysRevX.5.041036}, moving the system closer to the regime where the transmission dip on resonance is small.
	
	\begin{figure}
		\subfloat[SEM of a typical device.]{\includegraphics[scale=0.28]{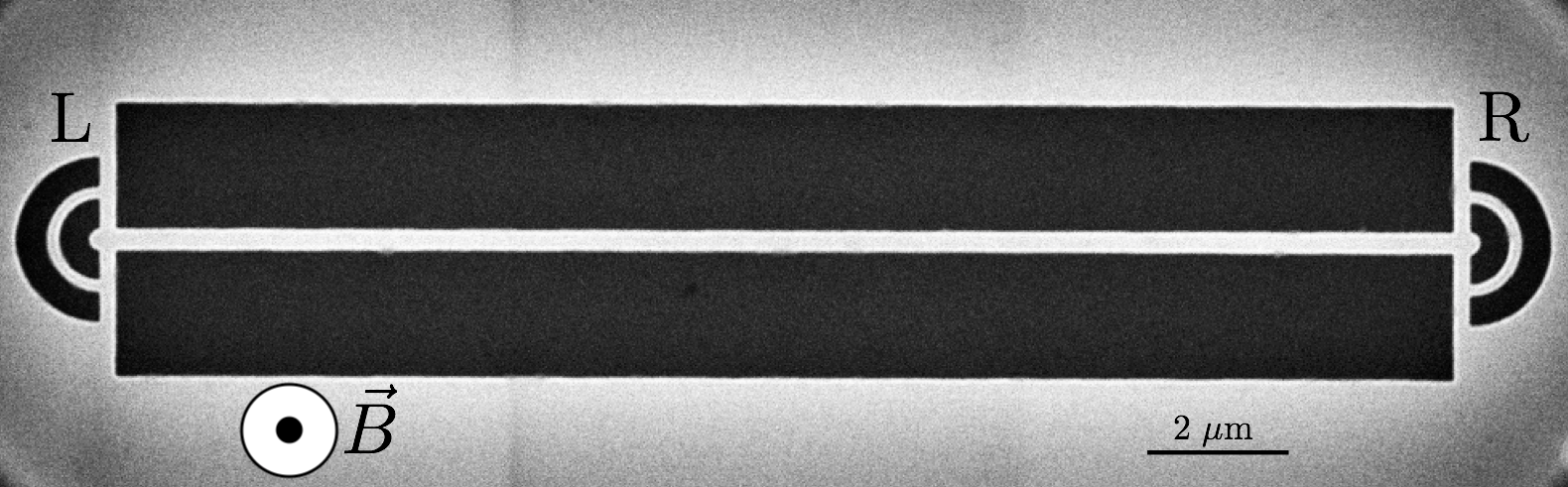} \label{fig:1a}} \\
		\subfloat[Selection.]{\includegraphics[scale=1]{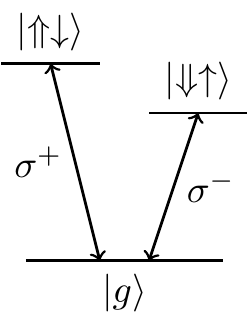} \label{fig:1b}} \hspace{11pt}
		\subfloat[Directionality.]{\includegraphics[scale=1]{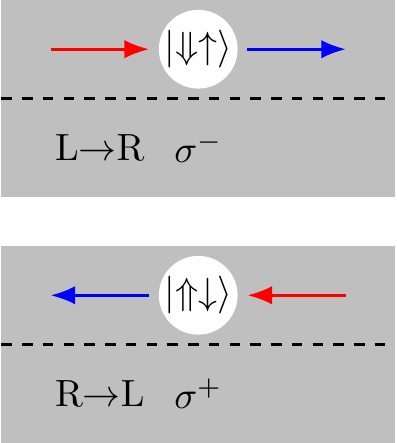} \label{fig:1c}} \\
		\subfloat[Photoluminescense.]{\includegraphics[scale=1]{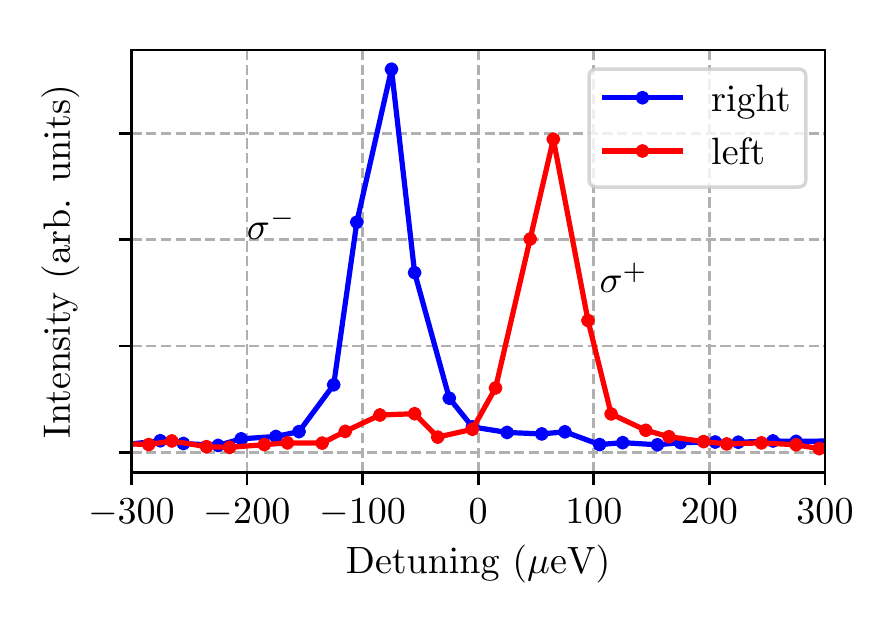} \label{fig:1d}}
		\caption{(a) Scanning electron microscope (SEM) image of a typical nanobeam waveguide with the left (L) and right (R) out-couplers labelled, along with the direction of the out of plane magnetic field $\vec B$. Such structures support both longitudinal and transverse field modes and subsequent propagation of in-plane circularly polarised light. (b) Selection rules for exciton transitions. The $\{ \Uparrow, \Downarrow \}$ and $\{ \uparrow, \downarrow \}$ symbols refer to hole and electron spins, respectively. The splitting of the transitions is caused by a Faraday-geometry magnetic field. (c) Directional spin-photon coupling. Emission and resonant scattering are shown by the blue and red arrows respectively. The exciton spin states that are coupled to the mode are indicated with the same notation as in (b). (d) PL Spectra for the Zeeman components of the chirally coupled QD at $B=1$~T, collected from the left and right outcouplers (see Fig.~1(a)). The observed linewidths are resolution limited and are also likely increased by the non-resonant excitation employed. The central energy (zero detuning) corresponds to $1.35743$~eV.}
		\label{fig:1}
	\end{figure}
	
The studies of non-reciprocal behaviour were carried out on a QD located at a chiral point (C-point) of a nanobeam waveguide, where opposite circular polarizations propagate in different directions~\cite{ncomms}. The structure consisted of a single self-assembled InGaAs quantum dot embedded within a single-mode, suspended vacuum-clad GaAs waveguide with out-couplers at its ends for efficient photon extraction, as shown in Fig.~\ref{fig:1a}. (See \textit{Methods} for further details of the sample.) The selection rules shown in Fig.~\ref{fig:1b} imply that opposite spin excitons couple to modes propagating in opposite directions. This applies both to emission, as shown schematically by the blue arrows in Fig.~\ref{fig:1c}, and to resonant scattering of incoming photons, as represented by the red arrows. 

\begin{figure}{\includegraphics[scale=1]{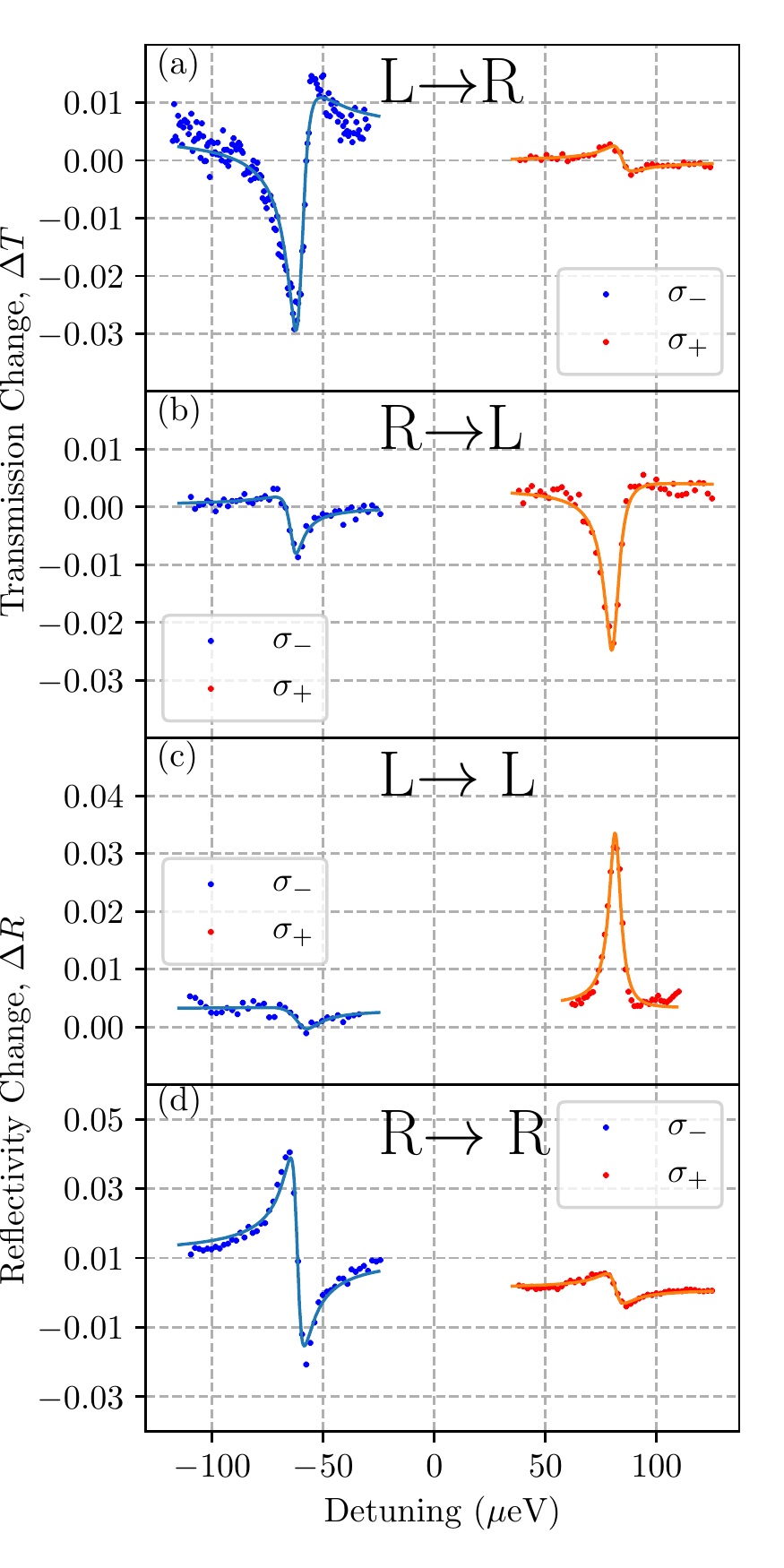}} 
	\caption{Differential transmission and reflectivity spectra for the chirally-coupled QD at $B = 1$~T: (a) transmission change $\Delta T$, L$\rightarrow$R (left to right) propagation; (b) transmission change $\Delta T$, R$\rightarrow$L propagation; (c) reflectivity change $\Delta R$, input from left; (d) reflectivity change $\Delta R$, input from right.  Differential spectra are used to isolate the resonant contribution of the QD transition. (See \textit{Methods}.) The black solid lines show the results of Fano lineshape fits according to Eq.~\ref{eq:defy}.}
	\label{fig:2}
\end{figure}

QDs near C-points were identified by exciting from above the waveguide with a non-resonant laser at 808\,nm and collecting the photoluminescence (PL) from the left and right out-couplers~\cite{ncomms,Coles_2017}. The PL spectra, with a magnetic field of $B=1$~T applied out of the waveguide plane, for the QD employed in this work are shown in Fig.~\ref{fig:1d}. Clear evidence of directional emission is present with $\sigma^+$ light propagating predominantly to the left and $\sigma^-$ predominantly to the right, as in Fig.~\ref{fig:1c}. The large degree of directionality shows the strong chiral coupling for this particular QD. The unidirectional emission contrast was calculated as in \cite{ncomms,Coles_2017} from the relative intensity of the Zeeman components $I^{\sigma +}$ and $I^{\sigma -}$  measured at a particular out-coupler:
\begin{align}
C = \frac{I^{\sigma +} - I^{\sigma -}}{I^{\sigma +} + I^{\sigma -}} \label{eq:eq1} . 
\end{align}
Over 50 randomly positioned QDs were examined to find those with high spin-dependent directionality. For the QD employed for the data in Fig.~\ref{fig:2}, the directional PL contrast ratios were  $C_\text{L}=0.84$  and $C_\text{R}=-0.91$ for the left and right out-couplers, showing the strongly chiral coupling for this particular QD.

Having identified a chirally-coupled QD, the non-reciprocal behaviour in resonant transmission was probed. A tunable single-frequency laser was input to one of the out-couplers and the transmitted light detected from the opposite out-coupler. An 808 nm non-resonant repump laser was applied to stabilise the QD charge state \cite{Makhonin}; no resonant transmission dips were observed without the repump laser. The QD charge state was not known with certainty but this was not important as, under the applied magnetic field of $B=1$~T, both charged and neutral excitons emit circularly-polarised light that couples to chiral fields \cite{ncomms}. In fact, it is most likely that we observed a charged exciton, since a repump laser creates free electron-hole pairs. 
The use of the repump laser permits the measurement of differential transmission and reflectivity spectra, where the contribution of the resonant QD transition under study is clearly identified. (See \textit{Methods}.)

Differential transmission spectra for L$\rightarrow$R propagation are shown in Fig.~\ref{fig:2}a and for the reverse case of R$\rightarrow$L propagation in Fig.~\ref{fig:2}b. Energies are measured as a function of detuning from the exciton transition energy at $B=0$~T. Clear transmission features from both the $\sigma^-$ and $\sigma^+$ exciton transitions are seen. However, on comparing Figs.~\ref{fig:2}a and ~\ref{fig:2}b, it is apparent that the $\sigma^-$ transition is dominant for L$\rightarrow$R propagation, whereas $\sigma^+$ is dominant for R$\rightarrow$L propagation, providing clear evidence for non-reciprocal behaviour in resonant transmission. All spectra were collected with an incident laser power of 50 nW and weak saturation of the QD exciton transition is occurring at this power. We discuss the saturation in detail and its effect on the spectra when we go on to model the system but note now that the maximum transmission dip is reduced from more than 3\% at lower powers to 2.5\%. 

The differential transmission spectra in Figs.~\ref{fig:2}a and \ref{fig:2}b have dispersive Fano-like lineshapes which arise from the interaction of the QD with the weak Fabry-P\'{e}rot cavity formed by reflections from the out-couplers of the sample  \cite{Javadi_2015}. The lineshape is determined by the phase difference between the QD optical response and continuum, and is highly sensitive to the wavelength of the transition and the precise position of the QD relative to the Fabry-Perot modes. This in turn depends on the propagation direction and the coupling of the incoming beam to the waveguide modes, a function of the measurement geometry. Positive signals occur when the QD resonance shifts the system to a point on the Fabry-P\'{e}rot mode with higher overall transmission, giving a larger increase in transmission than the drop caused by incoherent scattering. The contrast ratio was quantified by fitting the data to Fano line-shapes. (See \textit{Methods}). The fits give directional contrast ratios, defined by Eq.~\ref{eq:defC}, of $-0.86$ and $0.54$ for L$\rightarrow$R and R$\rightarrow$L propagation, respectively. On noting that L$\rightarrow$R propagation in transmission corresponds to R detection in PL, and vice versa for R$\rightarrow$L propagation, it is apparent that these contrasts correlate well with those obtained in PL, with the detailed differences likely originating from the different excitation regimes.
The asymmetry in the directionality between the two propagation directions was observed previously in PL experiments \cite{ncomms} and is likely related to the intrinsic structural asymmetry of the QD. 

Figures~\ref{fig:2}c and ~\ref{fig:2}d present results obtained in the L$\rightarrow$L and R$\rightarrow$R reflection geometries respectively. 
As for the transmission, the normalized differential signal $\Delta R$  is plotted --- see \textit{Methods}, Eq.~\ref{eq:Delta R} ---
leading to the possibility of both positive and negative changes in the reflectivity. 
In Fig.~\ref{fig:2}c, the resonant laser is incident from the left grating-coupler and the signal is detected in back-scattering from the same grating. In marked contrast to the transmission experiment with the laser incident from the left (Fig.~\ref{fig:2}a),	
a stronger peak is seen in reflectivity for $\sigma^+$, with only a weak feature at $\sigma^-$. 
The opposite is observed when the laser is incident from the right coupler, as shown in Fig.~\ref{fig:2}d.	
The contrast ratios deduced from Fano fits to the differential reflectivity are $0.83$ and $-0.73$ respectively for L$\rightarrow$L and R$\rightarrow$R propagation. (See \textit{Methods}.) The contrasts have opposite signs to those measured for the same direction of incidence in the transmission data.

As a control experiment, we repeated the measurements for a non-chiral system, where the QD is positioned close to the centre of the waveguide.  We find that in both the transmission and reflection geometry, similar magnitude spectral features are observed for both spin states (see Supporting Information, Section S1). This provides strong evidence that the non-reciprocal effects we observe here are indeed due to chiral-coupling between the QD and waveguide.

The difference between the behaviour in transmission and reflection for the chirally-coupled QD, with opposite spins dominating in the two cases, is, at first, rather surprising; one might naively expect that the QD transition coupled most strongly to the mode would show the largest signals in both transmission and reflection. This would certainly be true for a non-chirally-coupled QD, but it is not the expected behaviour for a chirally-coupled QD, as we now discuss.

\begin{figure}	 
	\subfloat[System.]{\includegraphics[scale=0.8]{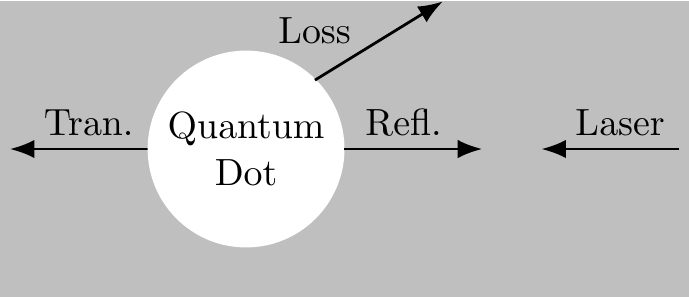} \label{fig:3a}} \\
	\subfloat[Transmission.]{\hspace{-1.2cm} \includegraphics[scale=0.8]{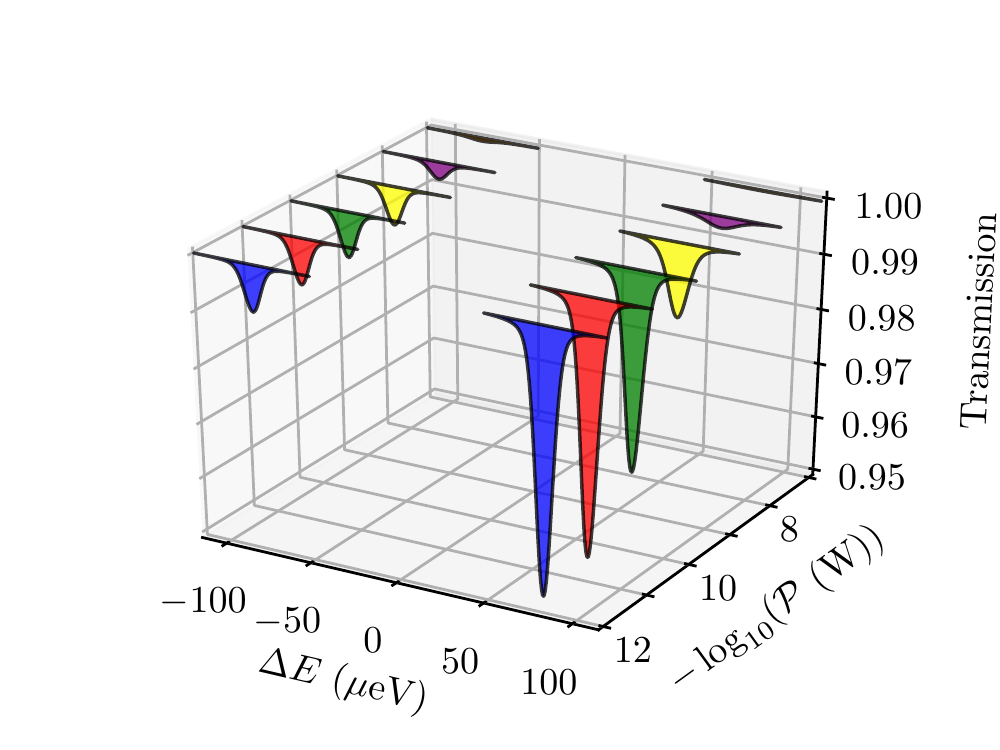} \label{fig:3b}} \\
	\subfloat[Reflection.]{\hspace{-1.2cm} \includegraphics[scale=0.8]{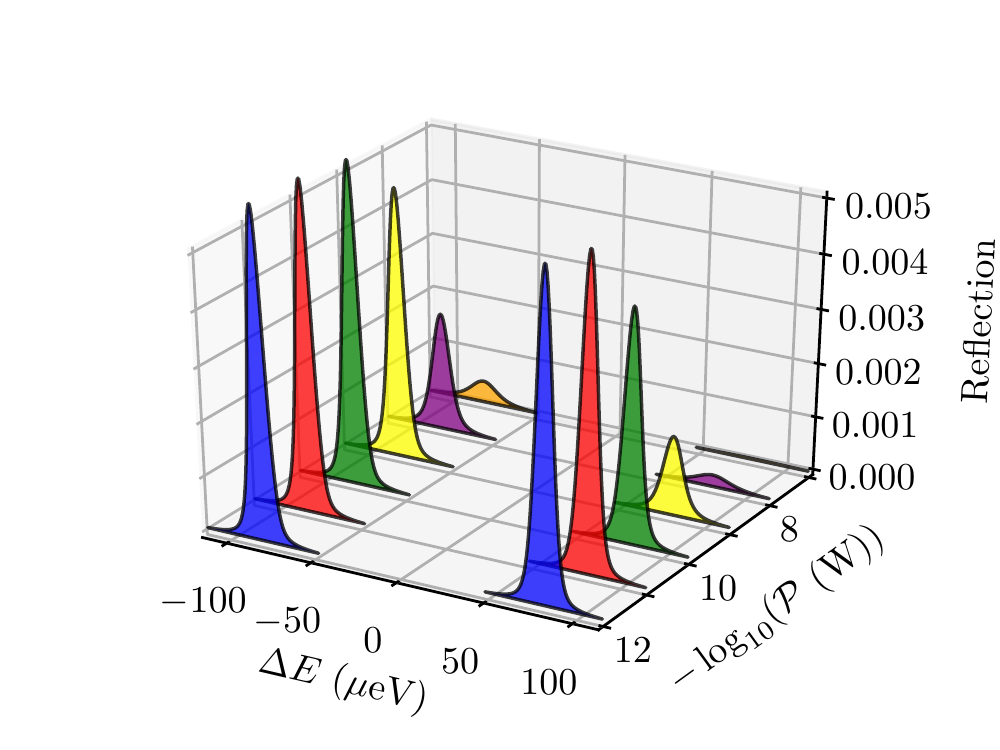} \label{fig:3c}}
	\caption{(a) Schematic of the system. A laser is coupled into a section of waveguide containing a QD, which is located at a C-point. The laser is either transmitted down the waveguide, reflected from the QD back down the waveguide or scattered into a continuum of free-space loss modes.
		(b)  Calculated transmission and  (c) reflection of the system for incident L$\rightarrow$R laser driving. Powers of 1, 10 and 100~pW are represented by blue, red and green curves respectively, with 1, 10 and 100~nW shown in yellow, purple and orange. The saturation data in the inset of Fig.~\ref{fig:4} indicates that the experimental conditions correspond to a power between 100~pW and 1~nW in the simulation.}
	\label{fig:model}
\end{figure}

The complete system under consideration is shown schematically in Fig.~\ref{fig:3a}. A QD is coupled to the single optical mode of a nanobeam waveguide and driven by a resonant laser field. The laser scatters from the QD and is either transmitted through the waveguide, reflected back in the direction of the laser input or lost from the sample.  The transmission of an ideal system with perfect directional coupling is 100\% for both QD spin states, but the behaviour of a realistic system is more complicated, being highly sensitive to a number of key parameters that account for the effects of imperfect directional coupling, an emitter-waveguide coupling ($\beta$-factor) less than unity, dephasing, spectral wandering and blinking.

In the Supplemental Information we model a system such as that shown in Fig.~\ref{fig:3a} using the well-known Input-Output formalism \cite{PhysRevA.30.1386}. The magnitude of the transmission reduction and reflection due to the QD is then calculated given knowledge of the QD-waveguide coupling, spectral wandering, blinking and dephasing time of system. In practice, we do not have access to these values directly and, as many of them contribute to the spectrum of the QD in the same manner (spectral wandering and pure dephasing for example), they cannot be deduced from the data. Furthermore, Fig.~\ref{fig:2} shows highly Fano-type behaviour, which originates from reflections at the input and output couplers.

Owing to the number of free parameters, it is then not possible to perform a first-principles fitting of the theory to the experimental data. We can however use good estimates for these parameters, derived from both experimental data and the literature, to show that the observed behaviour of the system is both reasonable and expected. For instance the coupling between the QD and waveguide is deduced from simulations \cite{ncomms} and the QD lifetime is directly measurable. We define the quantity $\beta_{\text d}$ as the fraction of the QD emission directed into the waveguide which propagates in R$\rightarrow$L direction and use a value of $\beta_{\text d}=0.95$. This implies a PL contrast ratio of $[\beta_{\text d} - (1-\beta_{\text d})] = 0.90 $, in agreement with the results in Fig.~\ref{fig:1d}. A lower limit of the pure dephasing time $\tau_{\text d}>120$\,ps is set by the $8\,\mu$eV QD linewidth, but the actual value of $\tau_{\text d}$ is longer due to the inhomogeneous broadening caused by spectral wandering. In the model we use $\gamma_{\text d} = (\tau_{\text d})^{-1} = (800\,\text{ps})^{-1}$ as a reasonable semi-quantitative estimate for a quantum dot in a nano-photonic environment under resonant excitation \cite{Javadi_2015,Sollner,Hallett:18,Makhonin}. The final parameters we require are estimates for the spectral wandering and blinking probability, $P_\text{dark}$. The spectral wandering is characterised by the parameter $\sigma$, the variance of the distribution, with $\sigma=4$~$\mu$eV giving a good fit to the measured 8~$\mu$eV QD linewidth. It is not possible to obtain a direct experimental estimate of $P_\text{dark}$ but previously reported values (e.g. in Refs.~\cite{Hallett:18,Javadi_2015}) fall within the range $0\leq P_\text{dark}\leq0.5$ and so we use $P_\text{dark}=0.25$ as a reasonable estimate. These parameters are summarised in Table~\ref{parameters}

\begin{table}[t]
	\begin{center}
		\begin{tabular}{l|c|c|l}
			Parameter & Symbol & Value & Notes\\
			\hline
			$ \beta$-factor & $\beta$ & 0.7 & calculated in ref. \cite{ncomms} \\
			Directionality & $\beta_{\text d}$& 0.95 & deduced from Fig.~\ref{fig:1d} \\
			Radiative lifetime & $\tau$ & 1 ns & 0.95~ns measured\\
			Dephasing time & $\tau_\text{d} $ & 0.8 ns & comparable to refs  \cite{Javadi_2015,Sollner,Hallett:18,Makhonin}\\
			Spectral wandering variance & $\sigma$ &  4 $\mu$eV & deduced from PL linewidth \\
			Dark probability &  $P_\text{dark}$ & 0.25 & within range of refs \cite{Hallett:18,Javadi_2015}\\
		\end{tabular}
	\end{center}
	\caption{Parameters used in the theoretical model.}
	\label{parameters}
\end{table}

The transmission spectra, calculated using the parameters of Table~\ref{parameters} are shown in Fig.~\ref{fig:3b}. The central energy of the QD is set at \protect{1.3~eV}, and the splitting between the low and high energy Zeeman components is \protect{0.16~meV} (as in the experimental data in Figs~\ref{fig:1d} and \ref{fig:2}), with the higher frequency component having the stronger coupling. The transmission dips are asymmetric, with the dip being stronger for the component preferentially coupled to the QD, in agreement with the experimental data and our intuitive understanding. The depth of the dips are close to those observed in Fig.~\ref{fig:2}, a maximum of 4\% experimentally and 5\% in the model, showing that the parameters used in the model are a reasonable approximation to the real system. We furthermore note that the size of the stronger dip is strongly dependent on the input power, which indicates that the system is saturated at powers of the order of $1$~nW impinging on the QD.

The qualitative behaviour in reflection is expected to be significantly different. Consider a R$\rightarrow$L input laser, coupling with relative efficiency of $\sim$95\% to the $\sigma^+$ dipole, which is in turn coupled with $\sim$5\% efficiency to the L$\rightarrow$R mode. By contrast, the $\sigma^-$ dipole couples with relative efficiency of 5\% to the R$\rightarrow$L mode but 95\% efficiency to the L$\rightarrow$R mode. As a first-order approximation and ignoring the interference effects that dominate in symmetrically-coupled systems, the fraction of the laser reflected into the L$\rightarrow$R mode is $\sim (95\% \times5\%)^2 \approx 0.2\%$ in \emph{both} cases. (Note that the reflected and transmitted intensities are dependent on the \emph{square} of the $\beta$-factor \cite{Thyrrestrup_2017}). This intuitive result with equal reflectivity peaks is reproduced by our numerical model provided that the power input to the system remains low, as shown in Fig.~\ref{fig:3c}. This low-power regime is characterised by the balancing of the stronger coupling to the laser with weaker back-scatter coupling, and vice versa. At higher powers asymmetry develops, as the more strongly coupled transition saturates first. 

\begin{figure}
	\includegraphics[scale=1]{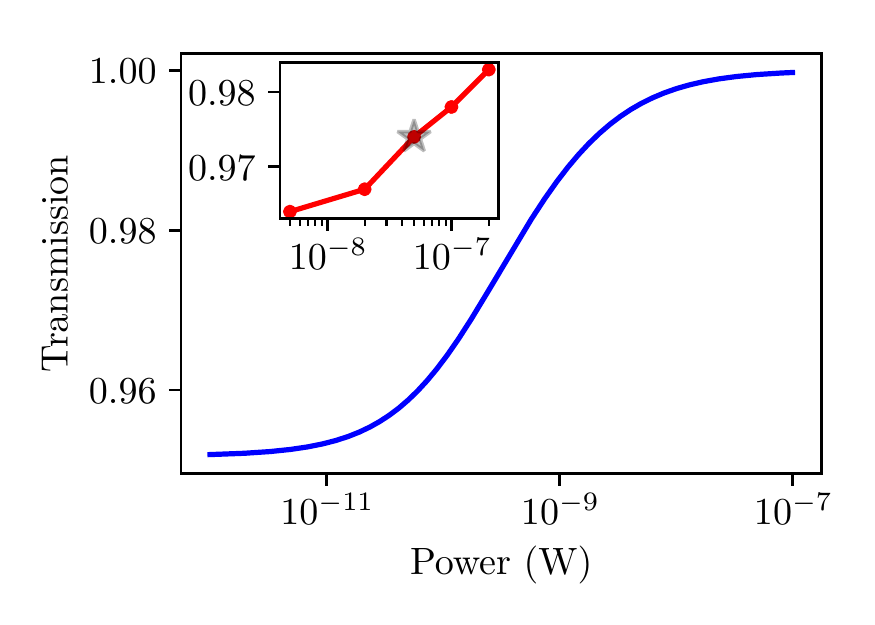}
	\caption{Theoretical power dependence of the main transmission dip on resonance for the preferentially coupled component. Inset: the experimentally measured dependence. The power employed for the resonant transmission and reflection experiments is indicated by the grey star marker. The spectra used to determine the experimental power dependence can be found in the Supplemental material.}
	\label{fig:4}
\end{figure}

In order to obtain a more thorough comparison of experiment and theory, we need to relate the power levels used in the model to those for the measured spectra. The powers used numerically are those \emph{within} the waveguide---after unknown coupling losses---and this makes direct comparison difficult. We can, however, calibrate the external power relative to that within the waveguide by analysing the predicted power dependence of the stronger transmission dip and comparing with experiment. In the main part of Fig.~\ref{fig:4}, we plot the power saturation dependence predicted by the model and show as an inset the experimentally determined power dependence. The spectra used to determine the experimental power dependence can be found in the Supplemental Information, Fig.~S2.

At low powers, below 10 pW, the main part of Fig.~\ref{fig:4} (the theory, the blue curve) confirms that the magnitude of the transmission dip is independent of incident laser power: fewer than one photon is interacting with the QD within its lifetime. As the power is increased up to 10 nW, the magnitude of the dip decreases as the QD can only interact with a certain fraction of the input light. At powers above 10 nW, the QD scatters an insignificant fraction of the incident photon flux and the fully saturated regime is entered. Experimentally we see very little reduction in transmission dip between 5 and 20nW and a marked reduction in transmission dip thereafter. By comparing points with the same 30\% reduction in transmission dip and cross-correlating, we are able to deduce that the power of 50 nW incident on the sample corresponds to a power of 100 pW to 1 nW within the waveguide. Having semiquantitatively calibrated the power, and returning to the theory curves of Fig.~\ref{fig:model}, we see that in this power range (represented by the green and yellow curves), the low frequency component still dominates in transmission, but the reflectivity has developed an asymmetry, with the higher frequency component being the stronger. The model thus qualitatively predicts the asymmetry in reflection observed in Figs~\ref{fig:2}c and \ref{fig:2}d through the different saturation powers for the two transitions. 

To take a specific example; in Fig.~\ref{fig:2}d we observe a contrast of -0.73 between the strongly and weakly coupled transitions. With the knowledge that the power incident on the QD lies in the range 100 pW to 1 nW, we now deduce this ratio from the reflectivity predictions of Fig.~\ref{fig:3c}. We see that for the green curve (100 pW), the contrast is -0.11 and -0.67 is predicted for the yellow curve (1 nW). The magnitude of experimental asymmetry in reflectivity is thus reproduced semi-quantitatively by the theory, providing good evidence for its origin in the direction-dependent saturation of the QD. Furthermore, we see that the green and yellow curves of Fig.~\ref{fig:3b} show that the transmission dips have contrasts of 0.5 and 0.33 for the more strongly and weakly coupled components respectively. This is in good agreement with the experimental data of Fig.~\ref{fig:2}b, which shows a ratio of 0.43.

Finally we note that a key parameter that can be calculated is the maximum phase shift, $ \Delta \phi$, that is imparted to a single photon as it is transmitted past the QD.  
The value of $\Delta \phi$ is $\pi$ for an ideal system with $ \beta \rightarrow 1$, $\beta_\text{d} \rightarrow 1$, and $(\tau_\text{d})^{-1} \rightarrow 0$.
Since the transmission probability of the ideal system is 100\%, a scalable quantum network can be implemented using this spin-dependent phase-shift\cite{Chiral-networks}.
In our system, $ \Delta \phi$ is calculated to be of the order of 0.4~rads if we ignore spectral wandering, which occurs on time scales longer than the emitter lifetime. 
The actual `useful' phase shift that could be extracted from an experimental sample would, of course, be lower, owing to spectral wandering and blinking.  
If we moved from a simple nanobeam to a photonic crystal platform, we could expect an increase in the $\beta$-factor from $\sim0.7$ to $\sim0.9$ and this would potentially boost $ \Delta \phi$ to $>0.6$~rads. The limiting factor at this point would then be the pure dephasing time, with a 3~ns time (as opposed to the 800~ps used for the modelling) giving $ \Delta \phi \sim 2$~rads. Since the dephasing time is an intrinsic property of the QD, a more realistic way to engineer this enhancement would instead be to reduce the radiative lifetime via a Purcell enhancement.

In conclusion, we have reported non-reciprocal transmission for a QD chirally coupled to the electromagnetic field supported by a nano-photonic waveguide. The key experimental result is the observation of a spin-dependent dip in the transmission spectrum, varying with the direction of propagation. The results observed in reflection geometry are initially counter-intuitive, with the more weakly-coupled transition giving a larger signal. We have shown that this is caused by partial saturation of the more-strongly coupled transition. We also show that the modelling of a realistic QD leads to a good understanding of the experimental data and what could be expected in non-ideal conditions. Further work with narrower-linewidth QDs in charge-stabilised structures \cite{Hallett:18,Thyrrestrup_2017} is expected to lead to the observation of deeper transmission dips down to $\sim 30\%$ limited by the $\beta$-factor, so that the power dependence of the reflectivity could be explored in more detail. Alternatively, the use of dots with Purcell enhancement \cite{Mahmoodian_2017}  and higher coherence could take us closer to the regime where a single photon can be deterministically imparted with a $\pi$-phase shift on transmission. 

The proof-of-principle results demonstrated in the paper have the potential to pave the way towards a spin-photon interface that would have applications in communication and quantum information technologies. For example, the use of QDs with high directionality but low $\beta$-factors could open the way to the realisation of on-chip, compact optical diodes operating at the single-photon level \cite{PhysRevX.5.041036}, or single-photon logic devices where the spin state is switched by external laser control \cite{Lodahl-review}, while moving to higher $\beta$ could lead to spin-based quantum networks \cite{Chiral-networks}, where quantum information is transmitted by emitted photons in a scalable, on-chip geometry. 

\section{Methods}
\textbf{\emph{Sample}}. The experiments were carried out on single QDs embedded in vacuum-clad single-mode waveguides. The InGaAs quantum dots were grown by the Stranski-Krastanov technique and were embedded in 140~nm thick GaAs regions, grown on top of a 1~$\mu$m thick AlGaAs sacrificial layer. Single-mode nanobeam waveguides of thickness 280~nm and height 140~nm were produced by a combination of electron-beam-lithography and wet and dry etching. Second-order Bragg-grating in/out-couplers \cite{Faraon,Luxmoore} were added on both ends of the waveguides for coupling to external laser fields. A scanning electron microscope image of a typical structure is shown in Fig.~\ref{fig:1a}. Further details of the sample structure and fabrication may be found in Ref.~\cite{Makhonin}.

\textbf{\emph{Experimental set-up}}. The measurements were made at 4~K in a confocal system with separate control of the excitation and detection spots. The spatial resolution was 1--2~$\mu$m \cite{Luxmoore-APL} and a Faraday-geometry magnetic field $B=1$ T was applied to split the $\sigma^+$ and $\sigma^-$ Zeeman transitions, as shown in Fig.~\ref{fig:1b}. This provided a convenient method to observe the interactions of a resonant laser field with well-defined spin states of the QDs within the mode-hop-free scan range of the laser. 

A weak non-resonant 808~nm repump laser with power 10~nW was used to stabilize the charge state of the dot \cite{Makhonin}. The repump laser beam was mechanically chopped at 500 Hz, and lock-in techniques were employed to maximise the signal to noise in the detection of the resonant laser transmitted to the out-coupler \cite{Nguyen_2012}. The normalized differential transmission spectrum $ \Delta T$ was obtained by finding the difference between the detected intensity with and without the repump laser:
\begin{equation}
\Delta T=\frac{(I^T_{ON}-I^T_{OFF})}{I^T_{OFF}} \, ,
\end{equation}
where $I^T_{ON}$ is the transmitted signal with the re-pump laser on and $I^T_{OFF}$ is the background signal with no re-pump laser.
This differential signal gives the contribution of the quantum dot transition that is resonant with the laser.
The differential reflectivity $ \Delta R$ was defined equivalently:
\begin{equation}
\label{eq:Delta R}
\Delta R=\frac{(I^R_{ON}-I^R_{OFF})}{I^T_{OFF}} \, ,
\end{equation}
where the superscript $R$ indicates that the reflected signal is measured.

\textbf{\emph{Fitting}}. The fitting of the transmission and reflection data was performed using Fano lineshapes described by the following equation:  
\begin{equation}
y(\omega)=y_{0}+A\frac{(q\Gamma+\omega-\omega_{0})^2}{\Gamma^2+(\omega-\omega_{0})^2}, 
\label{eq:defy}
\end{equation}
where $y_{0}$ is a background level, $A$ is the signal amplitude, $q$ is the Fano parameter, $\Gamma$ is the  line broadening, and $\omega_{0}$ is the resonant frequency. The contrast ratio for the directional differential transmission and reflectivity were then calculated from the appropriate fitted amplitudes according to:
\begin{equation}
\label{eq:defC}
C = \frac{A^{\sigma +} - A^{\sigma -}}{A^{\sigma +} + A^{\sigma -}} \, ,
\end{equation}
where $A^{\sigma +}$ and $A^{\sigma -}$ are the Fano amplitudes for the $\sigma^+$ and $\sigma^-$ Zeeman components at the out-coupler under study.

\section{Supporting Information}

The Supporting Information is available free of charge on the ACS Publications website at DOI: 

The Supporting Information includes: PL, resonant transmission, and resonant reflectivity data for a symmetrically coupled QD; input-output formalism and discussion of the effects of spectral wandering and blinking.

\section{Information}
The authors declare no competing financial interest. DLH and DMP contributed equally to the project and the manuscript was written with contributions from all authors. This work was funded by the EPSRC (UK) Programme Grants EP/J007544/1 and EP/N031776/1. DLH is supported by an EPSRC studentship.

\providecommand{\latin}[1]{#1}
\makeatletter
\providecommand{\doi}
{\begingroup\let\do\@makeother\dospecials
	\catcode`\{=1 \catcode`\}=2 \doi@aux}
\providecommand{\doi@aux}[1]{\endgroup\texttt{#1}}
\makeatother
\providecommand*\mcitethebibliography{\thebibliography}
\csname @ifundefined\endcsname{endmcitethebibliography}
{\let\endmcitethebibliography\endthebibliography}{}

\end{document}